\documentstyle[12pt]{article}
\newcommand{\be}{\begin{equation}}
\newcommand{\ee}{\end{equation}}
\begin{document}
\title{THEORY OF THE PHASE TRANSITION FROM A DISORDERED CUBIC CRYSTAL TO A GLASS}
\author{{\bf J. M. Y\'{a}\~{n}ez$^{\dagger}$, M. I.  Molina$^{\dagger}$ and D. C. Mattis$^{*}$}
\vspace{1 cm}
\and
$^{\dagger}$Facultad de Ciencias, Departamento de F\'{\i}sica, Universidad de Chile\\
Casilla 653, Santiago, Chile.\\
$^{*}$Department of Physics, JFB Bldg. \#201, 115 S. 1400 E\\
Salt Lake City, UT 84112-0830, U.S.A.}
\date{}
\maketitle
\baselineskip 18 pt
\begin{center}
{\bf Abstract}
\end{center}

\noindent
We calculate thermodynamic properties of a disordered model insulator, starting from
the ideal simple-cubic lattice ($g = 0$) and increasing the disorder parameter
$g$ to $\gg 1/2$. As in the earlier Einstein- and Debye- approximations, the ground
state energy is discontinuous at $g_{c} = 1/2$.
For $g<g_{c}$ the low-T heat-capacity $C \sim T^{3}$ whereas for
$g>g_{c}$, $C \sim T$. The van Hove singularities disappear at {\em any} finite magnitude
$g$ of the disorder. For $g>1/2$ we discover novel {\em fixed points} in the self-energy
and spectral density of this model glass.

\vspace{1cm}

\noindent
$^{\dagger}$email: jyanez@macul.ciencias.uchile.cl, mmolina@abello.dic.uchile.cl\\
$^{*}$email: mattis@physics.utah.edu
\vspace{2 cm}

\noindent
PACS number(s):\ \ 63.50.+x, 63.20.Dj, 63.20.Mt, 65.40.+g 
\newpage

The anomalous thermal properties observed in dielectric glasses\cite{zeller,pohl},
has prompted numerous theoretical and experimental studies on the subject\cite{dos}.
Arguably, the best well-known model consists of the phenomenological two-level states (TLS)
proposed by Anderson\cite{anderson} some time ago. Although the physical nature of
the TLS remains elusive the theory does predict linear specific heat at low-$T$ and
enhanced ultrasonic attenuation, both universal properties of glassy materials\cite{ultrasonic}.
An alternative microscopic theory entirely based on phonons was subsequently advanced
by one of the authors and his collaborators\cite{mattis}. It relies for its results
on the interplay between disorder, parametrized by a dimensionless $g$, and anharmonicity,
parametrized by a dimensionless $\xi$ (the Gr\"{u}neisen parameter) that is ultimately taken to $0$.
The solutions of two ``toy'' models, the one based on the unperturbed Einstein model of the
phonon spectrum in the ``crystalline'' phase and the other, on the Debye model\cite{mattis, phd},
both exhibited the following features: at a critical threshold $g_{c}$ the system undergoes a
transition from a ``disordered crystal'' phase with partial long-range order (LRO) to a ``glassy'' phase 
characterized by a linear specific heat and a divergent Debye-Waller exponent-i.e. by
a total loss of LRO\cite{phd}.

Here, for the first time, we have extended this type of model disorder {\em cum}
anharmonicity into the study of a bone-fide {\em crystal} and have obtained closed-form
solutions in the simple-cubic lattice. This approach is the most realistic so far, given that
initially the phonon density-of-states of the ideal crystal exhibits the van Hove
singularities (vHS) thought to be characteristic of a crystal with LRO. The results are
also much richer than in the toy models. In Fig. 1 we see that at low $T$ our calculated
specific heat is in better agreement with experiment than is the specific heat from
molecular dynamics.

Principally our findings are as follows: In the disordered crystalline phase, where
the vHS might be thought to weaken with increasing $g$ and ultimately disappear, instead
they soften and disappear {\em immediately} at any finite $g>0$, {\em despite} the persistence
of LRO and of a finite Debye-Waller exponent up until $g \ge g_{c}$. There is an exothermic
first-order phase transition as $g$ is increased above $g_{c} = 0.5$. Nevertheless the
elastic properties in the glassy phase are intimately related to those of the crystal, as we shall
show. Additionally,
in the glassy phase $g > g_{c}$, we discovered that the spectral density function $\rho$
exhibits {\em fixed points}: frequencies at which the spectral density is independent
of $g$. This feature bring us close to the ultimate goal of being able to express a microscopic
{\em law of corresponding states} for glassy materials. The specific heat exhibits a Debye
$T^{3}$ law throughout the crystalline phase, up to $g_{c}$. The exact coefficient $B(g)$ is given
in closed form below. Beyond $g_{c}$ the low-T specific heat is linear in $T$,
with coefficient $A(T)$ also given in closed form below.

\noindent
\paragraph{THE HAMILTONIAN.}\ \ In an effort to make this paper self-contained
we review here the main features of the model.
The starting point is the Hamiltonian of the reference crystal
\begin{equation}
\label{e:h0}
  H_{0} = \sum_{{\bf k}} \omega_{k} \left( a_{k}^{\dag} a_{k} 
  + \frac{1}{2} \right) \hspace{1cm}(\hbar=1)\, ,
\end{equation}
where ${\bf {k}}$ contains the wavevector and polarization of the phonons. The operator
$a_{k}^{\dag}$ ($a_{k}$) are the familiar creation (destruction) phonon 
operators.  Disorder is introduced in the model v\`{\i}a the rather general phonon
scattering term
\begin{equation}
\label{e:h1}
  H_{1} = \frac{1}{4} \sum_{{\bf k},{\bf  k'}} M({\bf k},{\bf k'}) \sqrt{\omega_{k}
  \omega_{k'}} Q_{k} Q_{k'} \, ,
\end{equation}
where $Q_{k}=a_{-k}^{\dag}+a_{k}$ and the $M(k,k')$ are random scattering 
amplitudes which depend on the location and type of disorder. We model $M(k,k')$
by a set of random phases $(g/\sqrt{N})\ \exp[i\ \theta(k,k')]$,
where the $\theta(k,k') = - \theta(k',k)$ are random and distributed uniformly in $[0, 2\pi]$ and
$g$ measures the strength of the disorder ($g^{2}$ is
proportional to the concentration of defects in the crystal). This choice is what allows the calculation
of the exact spectrum of eigenvalues of the random matrix\cite{note}.
Anharmonic effects
are introduced next in order to stabilize the spectrum, since $H_{0} + H_{1}$
possess an unstability threshold where the phonons become overcoupled\cite{mattis, phd}.
A simple quartic term suffices for this purpose,
\begin{equation}
\label{e:h2}
  \xi \frac{1}{N\omega_{0}} \sum_{k,k'} \omega_{k}\omega_{k'}
  Q_{k}Q_{-k}Q_{k'}Q_{-k'} \, ,
\end{equation}
where $\xi$ is the Gruneisen parameter, typically a small quantity, expressed here in
dimensionless form. $N$ is the number of
normal modes and $\omega_{0}$ is a characteristic frequency of the reference crystal.
Then a symmetry-breaking transformation $Q_{k} \rightarrow Q_{k}^{\dagger} + (f_{k}/\xi^{1/2})$ is
performed in order to maintain all the $\omega^{2}$'s positive. This shift, required
for $g>1/2$ to restore stability in the dynamics, produces
new terms linear in $Q$'s that have coefficients $O(1/\xi^{1/2})$ but that arise from
two distinct sources: from the quadratic terms (linear in the $f$'s) and from the quartic
(Gr\"{u}neisen) terms (third-order in the $f$'s). These must be canceled if the
individual atoms are all to be in positions of stable equilibrium. The relevant equations admit
only a trivial solution $f_{k} \equiv 0$ for $g < 1/2$, but for $g > 1/2$ the cancellation is
effected by choosing $f_{k}$ to be an eigenfunction of the random scattering matrix. (To minimize the
free energy one picks the eigenvalue belonging to the largest eigenvalue.) The ground state energy is then
a discontinuous function of $g$ at $g_{c} = 1/2$. At this value of the disorder the original locations of the atoms are no longer positions of stable equilibrium. Each atom moves to new, random, position that is essentially uncorrelated with the motion of the other atoms.
This picture is confirmed by a calculation of the Debye-Waller factor; we find it to be finite below
$g_{c}$ but divergent ($\propto 1/\xi$) once $g$ exceeds $g_{c}$, thereby confirming that all
long-range correlations are lost. The {\em dynamical} Hamiltonian\cite{mattis, phd}
governing lattice vibrations is, however continuous, and independent of $\xi$ in the limit. It is:

\begin{eqnarray}
\label{e:H}
  H & = & \sum_{{\bf k}}  \omega_{k}\left( a_{k}^{\dag}a_{k} + \frac{1}{2} \right)
  + \frac{g}{4}\sum_{{\bf k},{\bf k'}} \sqrt{\omega_{k}\omega_{k'}} M({\bf k},{\bf k'}) Q_{k}Q_{k'}  \nonumber \\
  &  & + \frac{1}{2}\left(g - \frac{1}{2} \right)\vartheta\left(g-\frac{1}{2}\right)
  \sum_{{\bf k}}  \omega_{k}
  Q_{k}Q_{-k} + O(\xi) +\nonumber\\
  	&   & + \theta\left( g - {1\over{2}} \right) \left\{ O(\xi^{1/2}) -
	O\left( \left( g - {1\over{2}} \right)^2/\xi\right)\right\}
\end{eqnarray}
where $\vartheta$ is the Heaviside step function. The origin of the individual powers in $\xi$ in $H$ is as
follows.

The $\xi$-independent correction to the elastic forces is the result of all contributions second-order
in the $f$'s in the quartic (Gr\"{u}neisen) term. The singular, discontinuous, shift ($\propto 1/\xi$) in
the ground state energy comes from the quartic term, when all four factors are $f$'s and from quadratic terms
in which both factors are $f$'s. But, however large
it may be, this shift in ground state energy does not affect the dynamics.

Two remaining contributions to the transformation vanish in the
lim $\xi\rightarrow 0$: terms with a single factor $f$'s in the
quartic term yield contributions are $O(\xi^{1/2})$ and those that are
quartic in $Q$'s (with no $f$'s) yield the original quartic term
$\propto \xi$.

The free energy of the model can be expressed as an integral over the coupling constant,
\be
F = k_{B} T \int_{0}^{\infty}\ d\omega [\ \rho_{0}(\omega) + \rho_{1}(\omega) + \rho_{2}(\omega)\ ]\ \log( 2\ \sinh(\beta \omega/2)),
\ \ \ \ \ \ \ \ \beta\equiv 1/k_{B} T\label{free}
\ee
where $\rho_{0}(\omega)$ is the density of states of the reference crystal,
\be
\rho_{0}(\omega) =
{2\over{\pi}}\ \omega\
\mbox{Im}\left[\ \int_{\small{FBZ}} \ {{d^{3} k\over{(2 \pi)^{3}}}}\ {1\over{\omega^{2} - \omega_{k}^{2}}}\ \right]\label{rho_0}
\ee 
and the densities of modes $\rho_{1}(\omega)$ and $\rho_{2}(\omega)$
are the contributions from disorder and anharmonicity\cite{phd}:
\be
\rho_{1}(\omega) = {2\over{\pi\ g^{2}}}\ \int_{0}^{1}\ {d\lambda\over{\lambda^{3}}}\ {\partial\over{\partial \omega}}
[\ -I(\omega,\lambda)\ R(\omega,\lambda)\ ] \label{rho1}
\ee
\be
\rho_{2}(\omega) = {\vartheta(g-1/2)\over{\pi\ g^{2}}}\ \int_{(2 g)^{-1}}^{1}\ {d\lambda\over{\lambda^{3}}}
 (2\lambda g -1)\ {\partial\over{\partial \omega}}
[\ -I(\omega,\lambda)\ ], \label{rho2}
\ee
and $R(\omega,\lambda)$, $I(\omega,\lambda)$ are the real and imaginary parts of the
the self energy $Z(\omega, \lambda)$ which obeys the transcendental equation\cite{mattis}
\begin{equation}
\label{e:Z}
  Z(\omega,\lambda) = \frac{(\lambda g)^{2}}{N}\sum_{k}\frac{\omega_{k}^{2}}{\omega^{2}-
  \omega_{k}^{2}(1+\phi+Z(\omega,\lambda))}\label{self}
\end{equation}
with $\phi = (2 \lambda g -1)\ \vartheta(\lambda g-1/2)$.

\noindent
\paragraph{RESULTS FOR THE SIMPLE-CUBIC LATTICE.}\ \ The integral (\ref{self}) can be performed analytically
if the dispersion relation are
$\omega_{k}^{2} = \omega_{0}^{2}( 3 - \cos(k_{1}) - \cos(k_{2}) - \cos(k_{3}))$. After
inserting this dispersion into Eq.(\ref{e:Z}) and taking the continuum limit, the equation for the self energy $Z(\omega, \lambda)$
can be written implicitly as,
\be
\label{e:ecZsc}
  W\left[3, \tau(\omega)\right] =   \\
  \left(1-3\frac{\omega_{0}^{2}}{\omega^{2}}(1+\phi+Z(\omega))\right)
  \left(1+\frac{Z(\omega)}{g^{2}}(1+\phi+Z(\omega))\right) \, , 
\ee
where the argument $\tau(\omega)$ is given by,
\begin{equation}
  \tau(\omega) = \frac{3(1+\phi+Z(\omega))}{3(1+\phi+Z(\omega)) - (\omega/
  \omega_{0})^{2}} \, 
\end{equation}
and $W[3,z]$ is the generalized Watson integral evaluated by Joyce\cite{joyce}, who expressed it
compactly in terms of complete elliptic integrals of the first kind\cite{AS72}:
\begin{equation}
\label{e:W}
  W[3, z] = \left(\frac{2}{\pi}\right)^{2} \frac{\sqrt{1-\frac{3}{4}
  x_{1}}}{1-x_{1}} K(k_{+}) K(k_{-}) \, ,
\end{equation}
where $x_{1} = (1/2) + (z^{2}/6) - (1/2)\sqrt{1-z^{2}}
\sqrt{1 - (z^{2}/9)}, x_{2} =  x_{1}/(x_{1} - 1),\newline
k_{\pm}^{2}  =  (1/2) \pm (1/4) x_{2} \sqrt{4-x_{2}} - (1/4)(2-x_{2})\sqrt{1-x_{2}}$. 
Figures 1 and 2 show the real and imaginary parts of $Z(\omega, g)$. The imaginary
part $I(\omega, g)$ shows the not unexpected systematic broadening of the bandwidth
with increasing disorder. $R(\omega, g)$ exhibits a change in curvature from convex to concave
at $\omega = 0$ as $g$ is increased from below to above $g_{c} = 0.5$. Correspondingly, the
slope of $I(\omega, g)$ vanishes at $\omega = 0$ for all $g < 1/2$ but acquires a finite value for
$g \ge 1/2$.

$Z(\omega, g)$ can be expanded at both the low and high frequencies. It is also seen to possess a
low-frequency fixed point in the glassy phase. From Eq.(\ref{self}), for $\lambda > 1$ and
$g > 1/2$  we deduce a sort of {\em duality} relation,
\be
Z(\omega, \lambda g) = \lambda\ Z(\omega/\sqrt{\lambda}, g)
\ee
connecting high frequencies, large disorder, to lower frequencies and smaller disorder.
$Z$ is inserted into the global density-of-states $\rho = \rho_{0} + \rho_{1} + \rho_{2}$
and the indicated integrals performed, with the results shown in Fig.3. For $g > 1/2$, {\em two}
fixed points in $\rho$ are discerned: one just below the characteristic frequency $\omega_{0}$ and the
other just below $3 \omega_{0}$. A new high-frequency ``tail'' grows beyond the second fixed point,
indicating an accumulation of spectral density at the highest frequancy as well as at the lowest.

It is also clear that the two van Hove {\em cusps} at $\omega/\omega_{0} = \sqrt{2}$ and $2$ are
sharp {\em only} at $g = 0$, and that $\rho$ losses its van Hove cusps and 
becomes analytic in $\omega$ for {\em any nonzero} value of the disorder, i.e.
already in the disordered crystal and not just
in the glassy phase! The theory works extraordinarily well for $g$ from zero into the glassy
phase at $0.5$ + ${\textstyle \epsilon}$, with almost every noncrystalline material being excellently characterized
at low $T$ by some ${\textstyle \epsilon} < 0.1$, and it remains trustworthy in every detail until $g$ exceeds $1$,
when the mean-field approximation to the quartic terms fails at high frequencies.

\noindent
\paragraph{THE SPECIFIC HEAT.}\ \ With a knowledge of $\rho$ the model's thermodynamic
properties, including specific heat, entropy and the other such functions are readily obtained.
For $g < 1/2$, we find at low $T$,
\be
c = {C(T)\over{N k_{B}}} = B(g)\ \ \left( {T\over{T_{0}}} \right)^{3}
\ee
where $B(g) = (4 \pi^{2}\sqrt{2}/5)\ (\ 1 + \gamma (\ [\ (1/2) +
\sqrt{(1/4) - g^{2}}\ ]^{-3/2} - 1 \ ) \ )$ and $\gamma \approx 1.43$. For $g > 1/2$,
\be
c = {C(T)\over{N k_{B}}} = A(g)\ \ \left( {T\over{T_{0}}} \right)
\ee
where $A(g) = \pi (2/3)^{5/2} \sqrt{W[3,1]} (1 - (2 g)^{-3/2} )
\approx 1.404 (1 - (2 g)^{-3/2} )$.

The high-$T$ specific heat is reasonably universal, always tending smoothly
to the Dulong-Petit limit because $\int_{0}^{\infty} d\omega\ \rho(\omega) = 3$.
Figure 4 illustrates the separate contributions $c_{0}, c_{1}$ and $c_{2}$ from the
crystal, disorder and anharmonicity respectively, at $g = 0.7$, over the
interval $0 < T/T_{0} < 0.6$.

\noindent
\paragraph{SPEED OF SOUND.}\ \  From our calculations using the present model
and on the earlier Debye model, we find that, at constant density,  the long
wavelength  speed of sound $s = (d\omega/dk)_{k\rightarrow 0}$, when expressed
as a ratio to the speed of sound  in the perfect crystal,
$v(g)\equiv s(g)/s(0)$, is a universal function of $g$ having a cusp minimum at
$g = 1/2$. Explicitly, we obtain:
\[
v(g) = \left\{ \begin{array}{ll}
	\left( {1\over{2}} + \sqrt{{1\over{4}} - g^{2}} \right)^{1/2} & \mbox{for $g <1/2$}\\
	g^{1/2} & \mbox{for $g > 1/2$}
		\end{array}
	\right.
\]
identical to Fig. 2 in Molina and Mattis [6] for the Debye model. The 
predicted drop of some 28\% as $g$ increased from $0$ to $1/2 + \epsilon$ should be 
confronted with experiment, after corrections for any changes in density with 
increasing disorder are taken into account.

We intend to publish the mathematical details of the various calculations elsewhere,
but in the meantime the interested reader may request them from the corresponding
author, M.I.M.
\vspace{2cm}

\centerline{ACKNOWLEDGMENTS}

\noindent
J.M.Y. was partially supported by FONDECYT grant 3980038. M.I.M. was partially
supported by the Millennium Science Initiative (Chile) under project p99-135-F

\newpage


\newpage

\centerline{{\bf Captions List}}
\vspace{2cm}

\noindent
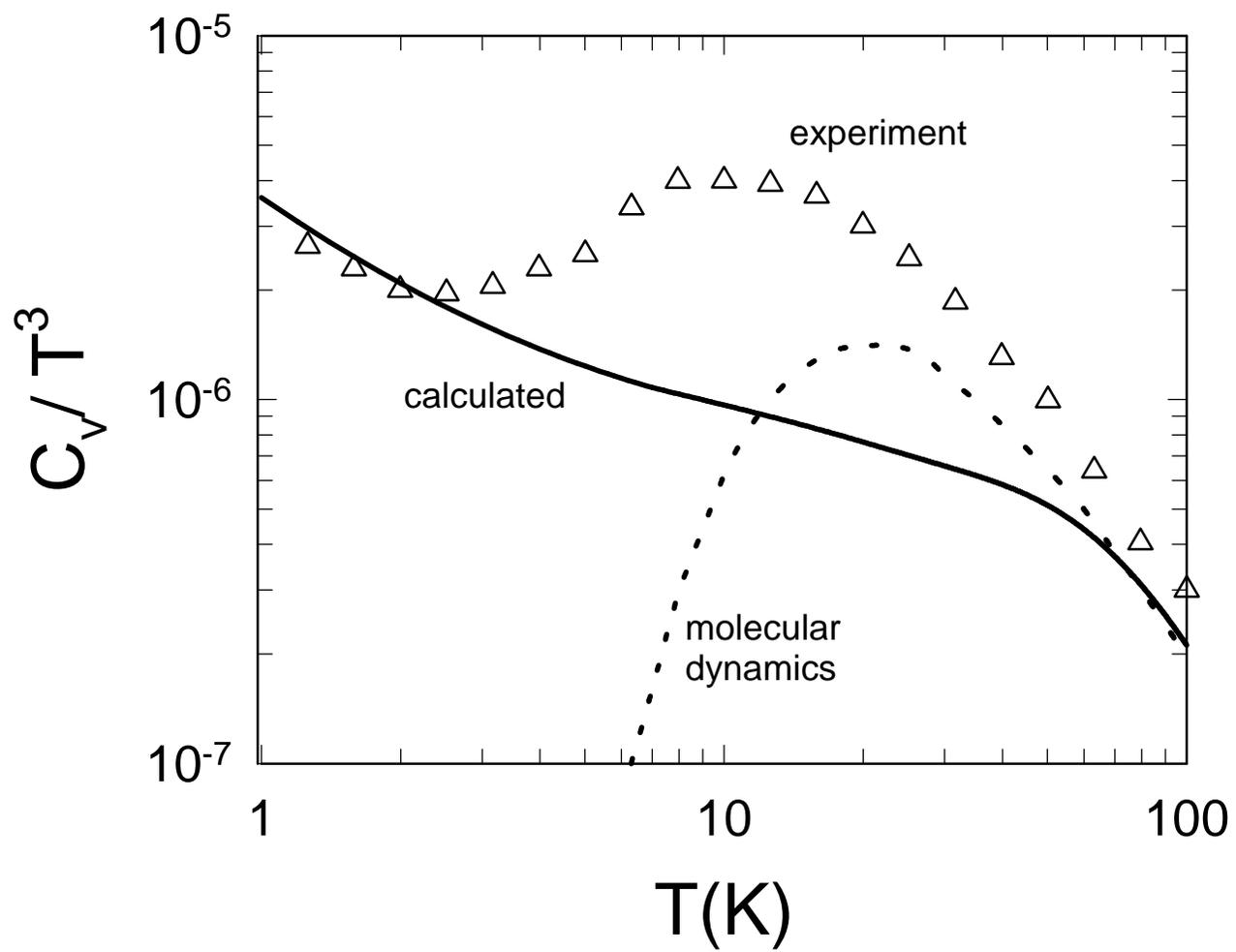
{\bf Figure 1:}\ \ Detailed comparison of specific heat of a-SiO$_{2}$ with
the present theory and with computer simulation molecular dynamics
(S. N. Taraskin and S. R. Elliott, Phys. Rev. B {\bf 56}, 8605 (1997)).
We observe excellent agreement of the present theory with experiment
at the low -and high- temperature end, although our result does lack
the ``bump'' at 10 K.
\vspace{0.4cm}

\noindent
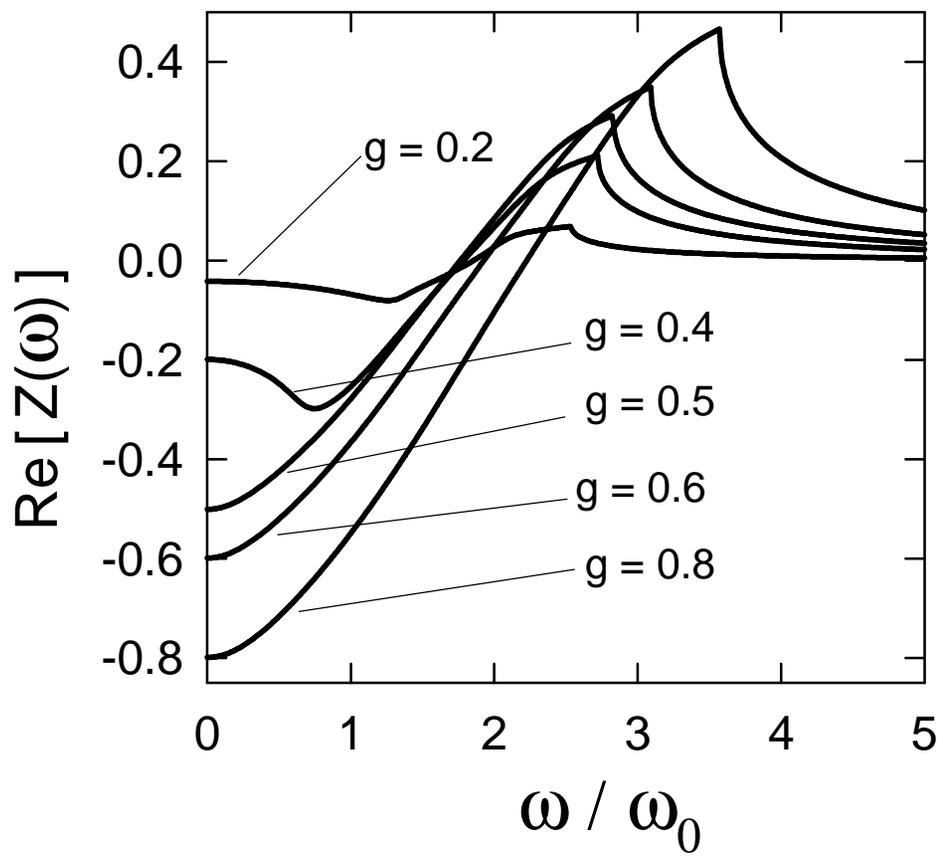
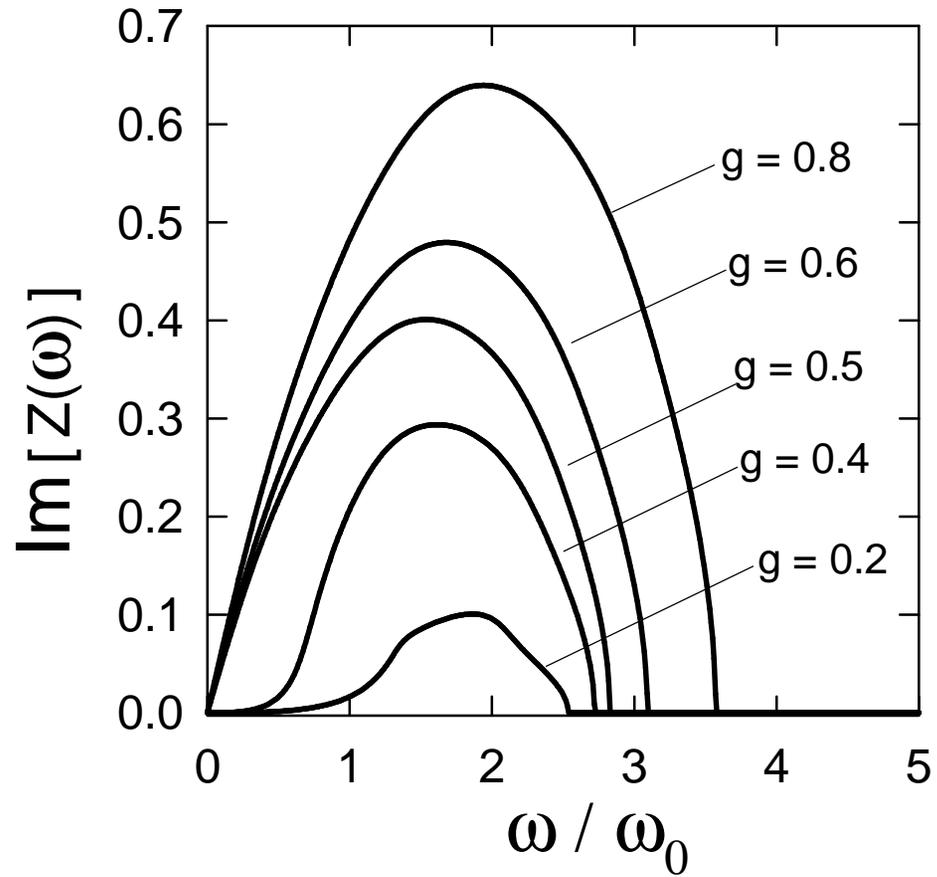
{\bf Figure 2:}\ \ Real and imaginary parts of the self-energy
$Z(\omega)$ as a function of frequency for several disorder
concentrations, below and above critical ($g = 1/2$). 
\vspace{0.4cm}

\noindent
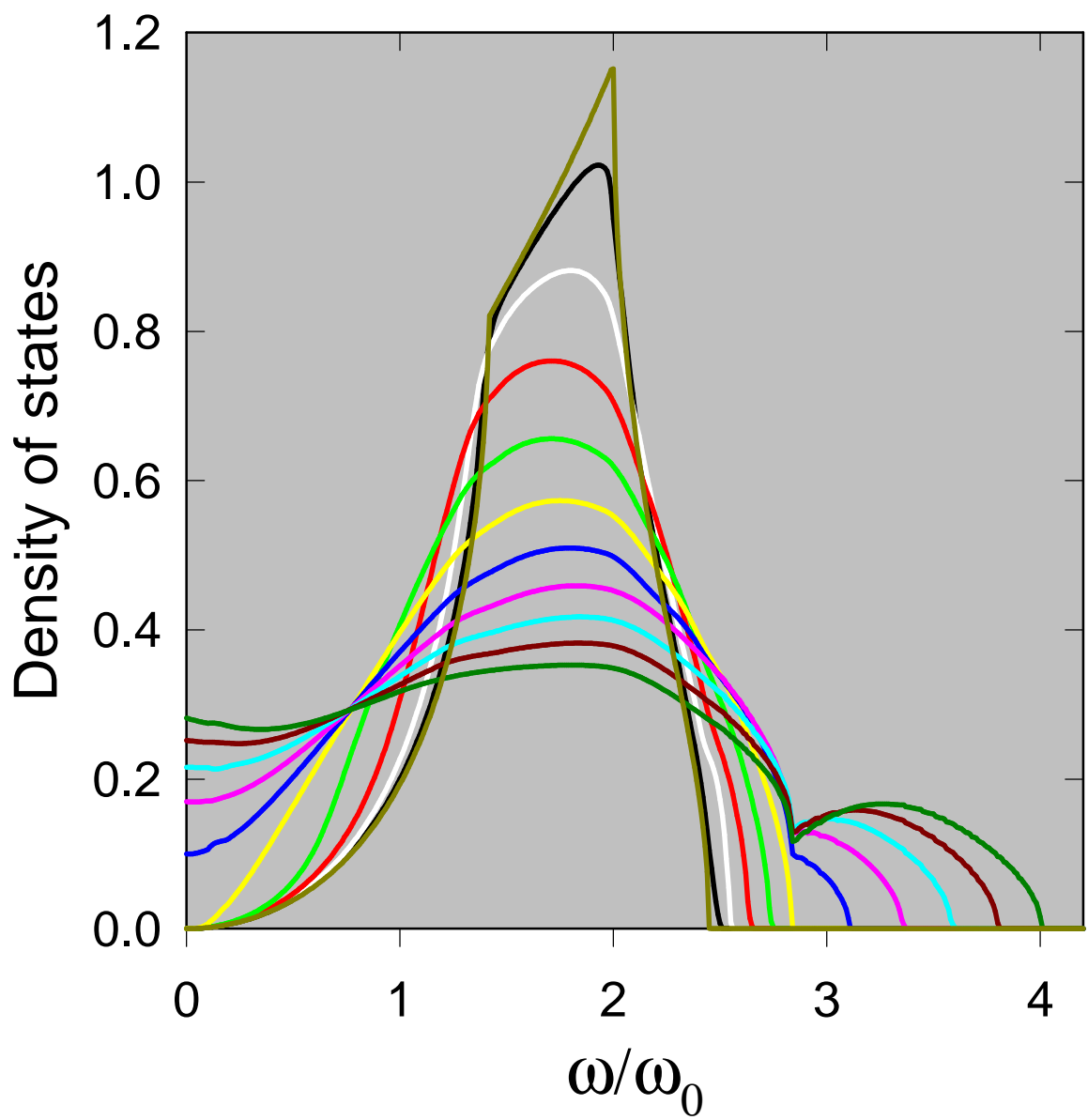
{\bf Figure 3:}\ \ Total density of states
$\rho(\omega) = \rho_{0}(\omega)+\rho_{1}(\omega)+\rho_{2}(\omega)$ as a function
of frequency for a wide range of disorder concentrations, proportional to $g^2$.
Arranged according to the height of the central maximum, we have
$g=0, 0.1, 0.2, ... $ down to $g =1$. 
\vspace{0.3cm}

\noindent
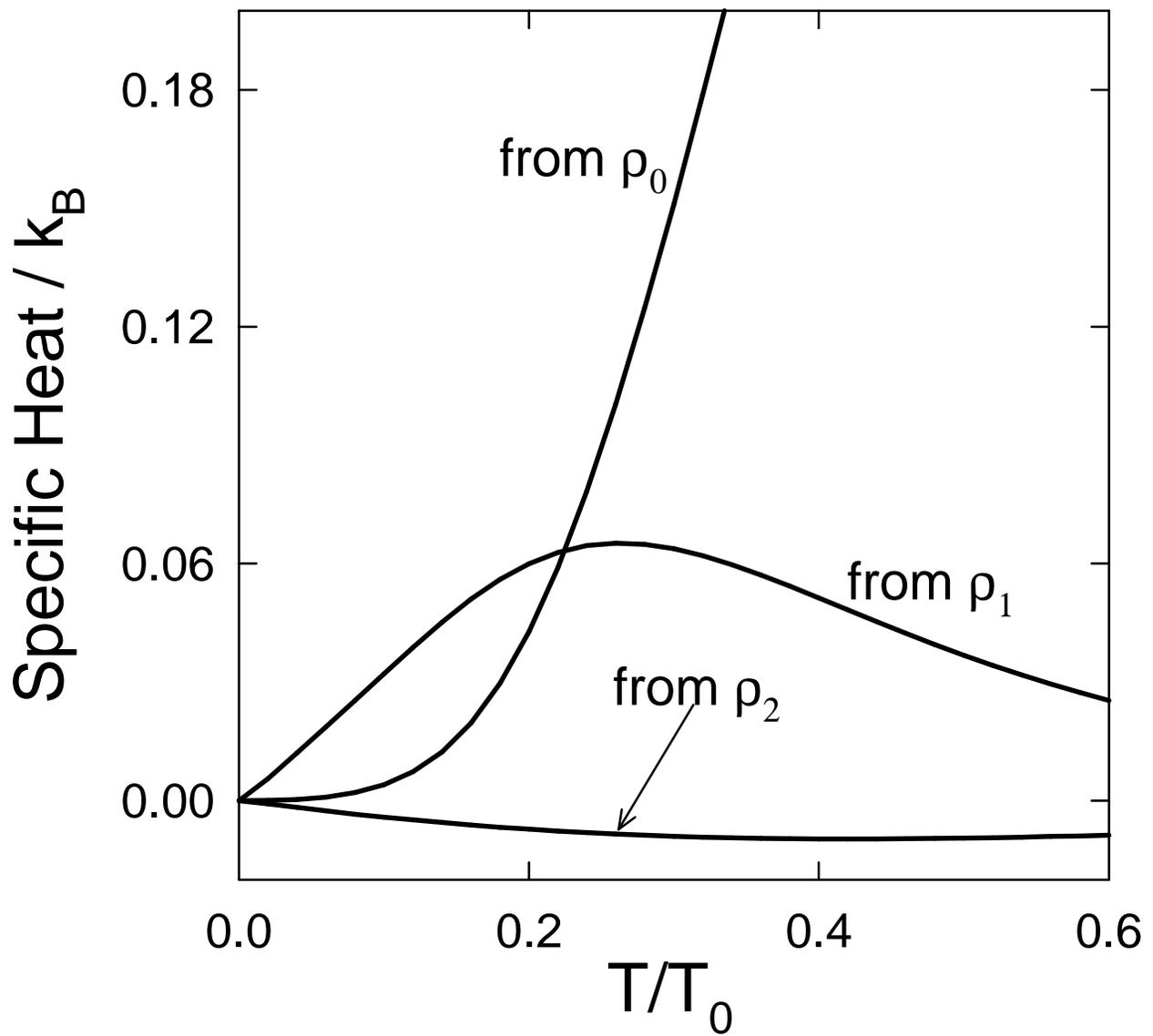
{\bf Figure 4:}\ \ The three distinct contributions to the low-temperature
specific heat from Eqs. (\ref{rho_0}), (\ref{rho1}) and (\ref{rho2}) in the glassy phase ($g = 0.7$).

\end{document}